\documentclass{article}
\usepackage{amsmath}
\usepackage{amssymb}
\usepackage{graphicx}

\begin{document}

\mbox{  }
\\[1,5 cm]
\mbox{  }

\begin{center}
\LARGE 
\textbf{Three-slit experiments}

\textbf{and quantum nonlocality}
\\[1,5 cm]
\normalsize
Gerd Niestegge
\\[0,3 cm]
\scriptsize
Fraunhofer ESK, Hansastrasse 32, 80686 Muenchen, Germany
\\[0,3 cm]
gerd.niestegge@esk.fraunhofer.de
\\[1,5 cm]
\normalsize
\end{center}
\normalfont
\itshape Abstract. \normalfont 
An interesting link between two very different physical aspects of quantum mechanics is revealed; 
these are the absence of third-order interference and Tsirelson's bound for the nonlocal correlations.
Considering multiple-slit experiments - not only 
the traditional configuration with two slits, but also configurations with three and more 
slits - Sorkin detected that third-order (and higher-order) interference is not possible in quantum mechanics.
The EPR experiments show that quantum mechanics involves nonlocal 
correlations which are demonstrated in a violation of the Bell or CHSH 
inequality, but are still limited by a bound discovered by Tsirelson.
It now turns out that Tsirelson's bound holds in a broad class of 
probabilistic theories provided that they rule out third-order interference.
A major characteristic of this class is the existence of a reasonable calculus of conditional probability
or, phrased more physically, of a reasonable model for the quantum measurement process.
\\[1,0 cm]
\\[1,0 cm]
\newpage
\section{Introduction}

The EPR experiments show that quantum mechanics involves nonlocal 
correlations which are demonstrated 
in a violation of the Bell inequalities \cite{ref-Bel} or CHSH 
inequality \cite{ref-CHSH} and have nowadays become the core of 
quantum information theory.
\\[0,3 cm]
However, the nonlocal quantum correlations are not unlimited; they obey a
bound found by Tsirelson \cite{ref-Tsi}. It follows algebraically from Hilbert 
space quantum mechanics without revealing any deeper physical or 
information theoretical principle behind it. The search for such 
a principle is a matter of ongoing research. 
Pawlowski et al. have recently introduced a promising new principle called information 
causality \cite{ref-Paw}. 
Janotta et al. consider a special
geometric property of the state space, known as strong self-duality \cite{ref-Jan}, and  
Navascues and Wunderlich study a principle which they name macroscopic locality \cite{ref-Nav}.
\\[0,3 cm]
Furthermore, Craig, Sorkin et al. consider a strongly positive joint decoherence functional in
a level 2 quantum measure theory \cite{ref-Craig}. A level 2 theory is characterized by an 
interesting property - the absence of third-order interference - which was discovered by
Sorkin \cite{ref-Sor}. Considering multiple-slit experiments - not only 
the traditional configuration with two slits, but also configurations with three and more 
slits - he detected that quantum mechanics rules out third-order interference.
The present paper also establishes a link between this property and Tsirelson's bound 
for the correlations, but uses a different framework and does not rely on a  
strongly positive decoherence functional.
\\[0,3 cm]
In the following sections, it will be shown that the absence of third-order interference implies the
validity of Tsirelson's bound for a broad class of probabilistic theories. 
A major characteristic of this class is the existence of a reasonable calculus of conditional probability
or, phrased more physically, of a reasonable model for the quantum measurement process \cite{ref-Nie0}.
In this class, the correlations  
cannot exceed Tsirelson's bound if
third-order interference is excluded or, vice versa,
a violation of Tsirelson's bound would imply the existence of third-order interference.
Mathematically, this is a quite simple consequence of the results from Ref. \cite{ref-Nie2}, 
but nevertheless provides an interesting link between two very different 
physical aspects of quantum mechanics. 
\\[0,3 cm]
The general framework for a probabilistic theory from Refs. \cite{ref-Nie0}, \cite{ref-Nie1} 
and \cite{ref-Nie2} and Sorkin's concept of third-order
interference \cite{ref-Sor} are briefly depicted in sections 2 and 3
before then turning to the major results in sections 4, 5 and 6. 
A general mathematical
inequality which follows from the absence of third-order interference is  
presented in section 4. Its relation to Tsirelson's bound is pointed out in 
section 5. This requires an understanding of the meaning of observables 
and compatibility in the general framework, 
which is elaborated in section 6. 

\section{The probabilistic framework}

In quantum mechanics, the measurable quantities of a physical 
system are re\-presented by observables. Most simple are those 
observables where only the two discrete values 0 and 1 are 
possible as measurement outcome; they are called \textit{events} or \textit{propositions}
and are elements of a mathematical structure called \textit{quantum logic}.
\\[0,3 cm]
Standard quantum mechanics uses a very special type of quantum logic; 
it consists of the self-adjoint projections on a 
Hilbert space or, more generally, of the self-adjoint 
projections in a von Neumann algebra.
\\[0,3 cm]
An abstractly defined quantum logic contains two specific elements $0$ and $\mathbb{I}$ 
and possesses an \textit{orthogonality relation},
an \textit{orthocomplementation}
and a \textit{partial sum operation} + which is defined 
only for orthogonal events \cite{ref-Nie0}, \cite{ref-Nie2}.
\\[0,3 cm]
The \textit{states} on a quantum logic are the analogue of the 
probability measures in classical probability theory, and 
\textit{conditional probabilities} can be defined similar to 
their classical prototype \cite{ref-Nie0}, \cite{ref-Nie2}. 
A state $\mu$ allocates the probability $ \mu(f) $ to each 
event $f$ (element of the quantum logic), and the conditional 
probability of an event $f$ under another event $e$ is the 
updated probability after the outcome of
a first measurement has been the event $e$; it is denoted 
by $ \mu(f \mid e) $.
\\[0,3 cm]
However, among the abstractly defined quantum logics, 
there are many where no states or no conditional 
probabilities exist, or where the conditional probabilities 
are ambiguous. Therefore, only those quantum logics where 
sufficiently many states and unique conditional 
probabilities exist can be considered a satisfying 
framework for general probabilistic theories. 
Such a quantum logic generates an order-unit space $A$
and can be embedded in the unit interval $\left[0,\mathbb{I}\right]$ 
$:=$ $\left\{a \in A : 0 \leq a \leq \mathbb{I} \right\}$ of this 
order-unit space; $\mathbb{I}$ becomes the order-unit \cite{ref-Nie2},
and each state on the quantum logic uniquely extends to a 
positive linear functional on $A$. The 
elements of $A$ are candidates for the observables.
\\[0,3 cm]
Not much knowledge of order-unit spaces is required for the 
understanding of the present paper and the interested reader
is referred to the monograph \cite{ref-Han}.
\newpage
\section{Third-order interference} 

Consider the following mathematical term $I_3$ 
for a triple of orthogonal events $ e_1 $, $ e_2 $ and $ e_3 $, 
a further event $f$ and a state $ \mu $:
\\
$$
\begin{array}{rcl}
  I_3 & := & \mu(f \mid e_1 + e_2 +e_3) \, \mu(e_1 + e_2 + e_3) \\
    &   &   \\
    &   & - \mu(f \mid e_1 + e_2) \, \mu(e_1 + e_2) \\
    &   &   \\
    &   & - \mu(f \mid e_1 + e_3) \, \mu(e_1 + e_3) \\
    &   &   \\
    &   & - \mu(f \mid e_2 + e_3) \, \mu(e_2 + e_3) \\
    &   &   \\
    &   & +  \mu(f \mid e_1) \, \mu(e_1) + \mu(f \mid e_2) \, \mu(e_2) + \mu(f \mid e_3) \, \mu(e_3) \\
\end{array}
$$
\\[0,1 cm]
This term $I_3$ was introduced by Sorkin \cite{ref-Sor},
and he detected that $ I_3 = 0 $ is universally valid in 
quantum mechanics. His original definition refers to probability measures 
on `sets of histories'. With the use of conditional probabilities,
it gets the above shape, which was seen by Ududec, Barnum and 
Emerson \cite{ref-Udu}. 
\\[0,3 cm]
For the three-slit set-up which Sorkin considered, the identity  $I_3 = 0$ 
means that the interference pattern observed with three open slits ($e_1 + e_2 + e_3$) is a simple 
combination of the patterns observed in the six different cases when only one 
or two among the three available slits are open ($e_1 , e_2, e_3, e_1 + e_2 , e_1 + e_3 ,$ and $e_2 + e_3$). 
Though Sorkin's theoretical discovery that this holds in quantum mechanics 
goes back to 1994, experimental testing has begun only recently and 
confirmed it to the accuracy achieved in the experiment \cite{ref-Sin}.  
\\[0,3 cm]
The new type of interference which is present whenever $ I_3 \neq 0 $ 
holds is called \textit{third-order interference}. 

\section {A mathematical inequality}

Quantum logics which do not exhibit third-order interference
(i.e., which satisfy the identity  $ I_3 = 0 $) have been studied 
in Ref. \cite{ref-Nie2}, and it has been shown that
there is a product operation  $ \Box $ in the order-unit 
space $A$ generated by the quantum logic, if the 
Hahn-Jordan decomposition property holds in addition (\cite{ref-Nie2} Lemma 10.2). 
The Hahn-Jordan decomposition 
for quantum logics is the analogue of the Hahn-Jordan decomposition for
signed measures in the classical case; it is a mathematical technical requirement
the details of which are beyond the scope of the present paper and can
be found in Ref. \cite{ref-Nie2}.
\\[0,3 cm]
The product $ a \Box b  $ is linear in $a$ as well as in $b$ and 
satisfies the inequality 
$\left\| a \Box b \right\|  \leq  \left\| a \right\| \: \left\| b \right\| $ ($a,b \in A$),
where $ \left\| \:  \right\| $ denotes the order-unit norm on $A$. 
The events $e$ become idempotent elements in $A$ (i.e., $e = e^{2} = e \Box e$), 
and $e \Box f = 0 $ for any orthogonal event pair $e$ and $f$.
Generally, however, the product is neither commutative nor associative nor power-associative.  
Moreover, the square $a^{2} = a \Box a$ of an element $a$ in $A$ need not be positive.
\\[0,3 cm]
If $a^{2} \geq 0$ holds for all $a \in A$, then
any $a_1, a_2, b_1, b_2 \in \left[-\mathbb{I},\mathbb{I} \right] $ := $ \left\{ a \in A: -\mathbb{I} \leq a \leq \mathbb{I} \right\}  $ = $ \left\{ a \in A: \left\| a \right\| \leq 1 \right\}  $ 
satisfy the following inequality:
\begin{equation}
\label{eq01}
\begin{array}{rcl}
-4 \sqrt{2} \mathbb{I} & \leq & a_1 \Box b_1 + b_1 \Box a_1 + a_1 \Box b_2 + b_2 \Box a_1 \\
& & \\
& & + a_2 \Box b_1 + b_1 \Box a_2 - a_2 \Box b_2 - b_2 \Box a_2 \\
& & \\
& \leq & 4 \sqrt{2} \: \mathbb{I}
\end{array}
\end{equation}
\textit{Proof}. $ 0 \leq ((1+\sqrt{2})(a_1 - b_1) + a_2 - b_2)^{2} + ((1+\sqrt{2})(a_1 - b_2) - a_2 - b_1)^{2} $ 
\begin{flushright}
$ + ((1+\sqrt{2})(a_2 - b_1) + a_1 + b_2)^{2} + ((1+\sqrt{2})(a_2 + b_2) - a_1 - b_1)^{2} = $
\end{flushright}
\begin{flushleft}
$ 4(2 + \sqrt{2}) (a_1^{2} + a_2^{2} +b_1^{2} +b_2^{2}) $
\end{flushleft}
\begin{flushright}
$ -4(1 + \sqrt{2}) (a_1 \Box b_1 + b_1 \Box a_1 + a_1 \Box b_2 + b_2 \Box a_1 
+ a_2 \Box b_1 + b_1 \Box a_2 - a_2 \Box b_2 - b_2 \Box a_2 ) $
\end{flushright}
The second $\leq$-sign in inequality (\ref{eq01}) now follows from 
$ a_1^{2} + a_2^{2} +b_1^{2} +b_2^{2} \leq 4 \: \mathbb{I} $
(note that $\left\|a_k^{2}\right\| \leq \left\|a_k\right\|^{2} \leq 1$ 
and thus $a_k^{2} \leq 1$ and, in the same way, $b_k^{2} \leq 1$ for $k=1,2$), 
the first one follows by exchanging $a_1$ with  $-a_1$ and $a_2$ with  $-a_2$.
\\[0,3 cm]
This is a simple transfer of Tsirelson's proof \cite{ref-Tsi} from 
quantum mechanics to the more general setting. However, this becomes 
possible only by using a deep result (Lemma 10.2) from Ref. \cite{ref-Nie2}. 
The meaning of the 
above inequality and its relation to Tsirelson's bound for quantum mechanical
correlations shall be studied in the following two sections.

\section {Tsirelson's bound}

The EPR experiments exhibit that quantum mechanics involves
stronger correlations between two 
spatially separated physical systems than possible in 
the classical case. Suppose that $a_1$ and $a_2$ are 
observables of the first system and $b_1$ and $b_2$
observables of the second system and that the spectrum of
each observable lies in the interval $[-1,1]$. 
Usually, it is assumed that $a_k$ and $b_l$ are compatible, i.e.,
$a_k$ commutes with $b_l$ ($k,l = 1,2$), 
but neither $a_1$ and $a_2$ nor $b_1$ and $b_2$ need commute.
The expectation values $c_{kl}$ of the products $a_k b_l$ ($k,l = 1,2$)
are a measure for the correlations
between the two systems then.
\\[0,3 cm]
If all four observables would commute with each other or if 
they were classical random variables, it would follow that
$$ 
\begin{array}{rcl}
\left| a_1 b_1 + a_1 b_2 + a_2 b_1 - a_2 b_2 \right| & \leq  & \left| a_1 \right| \: \left|b_1 + b_2\right| + \left| a_2 \right| \: \left|b_1 - b_2 \right| \\
& & \\
& \leq & \left|b_1 + b_2\right| + \left|b_1 - b_2 \right| \\
& &   \\
& \leq & 2 \\
\end{array}
$$
(note that commuting observables can be considered as functions) and therefore 
$$ \left| c_{11} + c_{12} + c_{21} - c_{22}  \right| \leq 2 $$
This is the CHSH inequality \cite{ref-CHSH}.
However, if neither $a_1$ and $a_2$ nor $b_1$ and $b_2$ commute, 
$c_{11} + c_{12} + c_{21} - c_{22} $ can exceed the value 2 and 
can reach the value $2 \sqrt{2}$ in certain EPR experiments. This shows
that quantum correlations do not obey the same rules as 
classical correlations. 
\\[0,3 cm]
Tsirelson \cite{ref-Tsi} discovered that $2 \sqrt{2}$ 
is the largest possible value for $c_{11} + c_{12} + c_{21} - c_{22}$
in quantum mechanics. Inequality (\ref{eq01}) now shows that
Tsirelson's bound still holds in a more general theory. It remains valid
if the theory includes a reasonable calculus of conditional probability,
rules out third-order interference, satisfies the 
Hahn-Jordan property and if squares in $A$ are positive. 
However, this requires an understanding of the meaning of observables 
and compatibility in the general framework, which shall be
discussed in the following section.

\section {Observables and compatibility}
 
Suppose that third-order interference is ruled out, that the Hahn-Jordan property 
holds and that squares are positive in the order-unit space $A$.  
Is each element in $A$ an observable then? An observable
represents a measurable physical quantity and should 
lie in an associative commutative algebra, ideally in an algebra of real
functions. 
\\[0,3 cm]
It can indeed be shown that any
associative commutative closed subalgebra of $A$ containing the order-unit $\mathbb{I}$
is an associative JB-algebra (JB-algebras are the Jordan analogue of 
the $C^{*}$-algebras; see \cite{ref-Han}) and thus isomorphic to the algebra of continuous real
functions on some compact Hausdorff space; this can be proved in the same way as
a very similar result in Ref. \cite{ref-Nie1}. 
Therefore, only those elements of $A$
which generate an associative subalgebra represent observables. 
\\[0,3 cm]
Simple examples
are the elements having the shape $ \alpha_1 e_1 + \alpha_2 e_2 + ... + \alpha_n e_n $
with mutually orthogonal events $e_1 , e_2 , ..., e_n $ and real 
numbers $\alpha_1 , \alpha_2 , ..., \alpha_n$.
Note that $ e_k \Box e_l = 0 $ for $k \neq l$ because of the orthogonality 
and that $ e_k \Box e_k = e_k $.
\\[0,3 cm]
Two observables are \textit{compatible} if they lie in a joint associative commutative 
subalgebra of $A$. The pair then behaves like two classical random variables and they
are simultaneously measurable. This represents the strongest level in the hierarchy 
of different compatibility and comeasurability levels studied in 
Ref. \cite{ref-Nie3} for events (not for observables). There, 
this strong level was named `\textit{algebraic compatibility}.' 
Since the distinction among the different levels is not needed in the present paper, 
the tag `algebraic' is dropped here.
\\[2 cm]
A state is a positive linear functional $\rho$ on the order-unit space $A$ with 
$\rho (\mathbb{I}) = 1$, and $\rho(a)$ is the expectation value of an observable $a$ in $A$.
For a compatible observable pair $a$ and $b$, $ a \Box b =  b \Box a$ holds, the product itself
is an observable and its expectation value $\rho(a\Box b)$ is a measure for the correlation of $a$ and $b$.
\\[0,3 cm]
With four observables $a_1, a_2, b_1, b_2 \in \left[-\mathbb{I},\mathbb{I} \right] $ such that 
$a_k$ and $b_l$ are compatible ($k,l = 1,2$), inequality (\ref{eq01}) yields that
$$ \left| c_{11} + c_{12} + c_{21} - c_{22}  \right| \leq 2 \sqrt{2} $$
holds for the correlations $c_{kl} = \rho (a_k \Box b_l) = \rho (b_l \Box a_k) $ ($k,l=1,2$). 
This means that Tsirelson's bound remains valid in this setting, although it is more general 
than quantum mechanics.
 
\section {Conclusion}

Using mathematical methods, an interesting link between two very different 
physical aspects of quantum mechanics has been revealed. It has been seen 
that Tsirelson's bound holds for the correlations 
in any probabilistic theory if third-order interference is ruled out and
three further assumptions are met; these are 
a reasonable calculus of conditional probability or, 
phrased more physically, a reasonable model for the quantum measurement process, 
the Hahn-Jordan decomposition property, which is a technical 
mathematical requirement, and 
the positivity of squares of the elements in the order-unit space $A$.
\\[0,3 cm]
If it is assumed that each element in $A$ is an observable, $A$ becomes a so-called JBW-algebra
(the Jordan algebra analogue of a $W^{*}$-algebra or von Neumann algebra; see \cite{ref-Han}). 
This is the major result (Theorem 11.1) in Ref. \cite{ref-Nie2}. However, this assumption
is not required for the derivation of Tsirelson's bound.
It has thus been seen that the validity of Tsirelson's bound is not restricted to JBW-algebras
and particularly not to Hilbert space quantum mechanics.
The exceptional Jordan algebra (consisting
of the self-adjoint $ 3\times 3$ matrices the entries of which are octonions) 
is a concrete example of a JBW algebra which is not included in Hilbert space quantum mechanics, 
but still satisfies Tsirelson's bound.
\\[0,3 cm]
If not all elements in $A$ are observables, the sum of two observables is an
element in $A$, but need not be an observable, and there is no 
linear structure for the observables anymore. However, is it so natural to 
postulate a linear structure for the observables?
The observables in section 6 as well as those usually considered in quantum mechanics
are real-valued observables. It is not very common to study complex-valued or
vector-valued observables, but such a postulate would be equally natural for them, 
however, is not even valid in Hilbert space quantum mechanics.
The usual real-valued observables are the self-adjoint operators ($ a = a^{*} $) 
which form a real-linear structure; the complex-valued observables 
are the normal operators ($ a a^{*} = a^{*} a $), but the sum of two normal operators
is not normal unless they commute. 
One might therefore imagine a more general physical theory 
with no linear structure for the observables - not even for the usual real-valued ones.
\\[0,3 cm]
Moreover, an interesting question is how the correlations would behave when 
the assumption that the squares of elements in the order-unit 
space $A$ are positive is dropped
or, perhaps more important, when third-order interference is permitted.
It may not be expected that Tsirelson's bound remains valid. So the question is whether 
$c_{11} + c_{12} + c_{21} - c_{22}$ can then reach the value 4 
(which is the algebraic maximum because $-1 \leq c_{11},c_{12},c_{21},c_{22} \leq 1$), 
or whether there are other bounds between the two in these cases. 
An example where the algebraic maximum 4 is reached 
was found by Popescu and Rohrlich \cite{ref-PR}; 
it can be implemented in a simple quantum logic, 
but it is not known whether this is possible
in a quantum logic with a reasonable calculus of unique conditional probability.
\\[0,3 cm]
The link between third-order interference and Tsirelson's bound has also been seen by
Craig, Sorkin and al. \cite{ref-Craig}. Their assumptions include the existence of a strongly 
positive decoherence functional. It might be possible to derive
the real part of the decoherence functional from the product operation $\Box$, but its 
imaginary part and particularly its strong positivity are not available in the more
general setting of the present paper.

\end{document}